\documentclass[12pt]{iopart}
\usepackage{graphicx}
\usepackage{cite}

\usepackage{iopams}  
\begin{document}

\title[Non-stationary $1/f$ noise in quantum dots]{$1/f$ noise for intermittent quantum dots exhibits 
non-stationarity and critical exponents}

\author{Sanaz Sadegh$^1$,Eli Barkai$^2$,Diego Krapf$^{1,3}$}

\address{$^1$ Department of Electrical and Computer Engineering, 
Colorado State University, Fort Collins, CO 80523, USA}
\address{$^2$ Department of Physics, Institute of Nanotechnology and Advanced Materials, 
Bar Ilan University, Ramat-Gan 52900, Israel}
\address{$^3$ School of Biomedical Engineering, Colorado State University, Fort Collins, CO 80523, USA}

\ead{krapf@engr.colostate.edu, eli.barkai@biu.ac.il}
\vspace{10pt}
\begin{indented}
\item[]August 2014
\end{indented}

\begin{abstract}
The power spectrum of quantum dot fluorescence exhibits $1/f^\beta$ noise, 
related to the intermittency of these nanosystems. 
As in other systems exhibiting $1/f$ noise, this power spectrum is 
not integrable at low frequencies, which appears to imply infinite 
total power. 
We report measurements of individual quantum dots that address this long-standing paradox. 
We find that the level of $1/f^\beta$ noise decays with the observation time. 
The change of the spectrum with time places a bound on the total power.   
These observations are in stark contrast with most measurements of noise in macroscopic systems 
which do not exhibit any evidence for non-stationarity. 
We show that the traditional description of the power spectrum with a single exponent $\beta$ 
is incomplete and three additional critical exponents 
characterize the dependence on experimental time. 
\end{abstract}


%
\vspace{2pc}

\submitto{\NJP}
%
%
%

\section{Introduction}
The power spectrum of many natural signals exhibits $1/f$ noise 
at low frequencies \cite{Hooge,eliazar2009}. This noise appears in an extremely broad range of systems that includes 
electrical signals in vacuum tubes, semiconductor devices, and metal films 
\cite{Mandelbrot,Dutta1981}, 
as well as earthquakes \cite{sornette1989}, network traffic \cite{csabai19941}, 
evolution \cite{halley1996ecology},  
and human cognition \cite{gilden19951}.  
All these systems are characterized by a power spectrum 
of the universal form $S(f) \sim { A/f^{\beta} }$, 
where the exponent $\beta$ is between $0$ and $2$ 
\cite{Mandelbrot,Keshner,Kogan}. 
The long time that has passed from the first 
discovery of this phenomenon \cite{johnson1925} led to multiple theories, 
competing schools of thought 
and many unresolved problems. 
One of the major problems lies in the fact that the spectrum is not integrable 
at low frequencies if $\beta>1$, i.e., $\int_0 ^\infty S(f) {\rm d}  f=\infty$. 
This is a paradoxical issue since the total power cannot be infinite, 
as implied by the divergence of the integral to infinity. 
In order to solve this paradox, Mandelbrot suggested 
that $1/f$ noises are related to non-stationary processes \cite{Mandelbrot}. 
However, $1/f$ noise in macroscopic systems, where a large number of subunits are intrinsically averaged, 
do not exhibit evidence of non-stationarity \cite{Kogan}, hence this famous paradox remains open. 
Moreover, these ideas have been often contested, for example, by assuming that there exist some 
low-frequency cutoff $f_c$ under which the non-integrable
$1/f$ spectrum is no longer observed. For this reason, 
several groups have increased the measurement time in an attempt to find this 
evasive low-frequency cutoff. For example, spectral estimations have been 
obtained for one cycle in three weeks in operational amplifiers \cite{Caloyannides} and one cycle in 
300 years in weather data \cite{mandelbrot1969some}. 
Despite such long measurements, no low-frequency cutoff was found in these systems.

During the last two decades, experimental work  
has shown that $1/f$ noise is also observed in a vast array of 
nanoscale systems. For example, such noise was observed in 
individual ion channel conductivity \cite{bezrukov2000,siwy2002origin}, 
electrochemical signals in nanoscale electrodes \cite{Krapf}, 
biorecognition processes leading to the formation of a complex \cite{Bizzarri},
and graphene devices \cite{Balandin}.
Noise in nanoscale systems is particularly intriguing due to their sensitivity to environmental
conditions. Furthermore, the characterization of noise properties in nanomaterials is an important challenge 
with direct applications in the stabilization of these materials for nanotechnology devices. 

A well investigated but still poorly understood case is
blinking in nanocrystals. These systems exhibit intermittency, namely
random switching between dark and bright states, 
with sojourn times distributed according to power laws with heavy tails \cite{Kuno,Shimizu,Frantsuzov}. 
This power-law behavior was shown by Brokmann et al. \cite{Brokmann} to induce unusual 
phenomena such as ergodicity breaking and non-stationary correlation functions, 
which are discussed here in the summary.  
Physical models underlying a power-law sojourn time distribution are based on distributed
tunnelling mechanisms or diffusion controlled reactions \cite{Shimizu,Frantsuzov2008,Stefani}.
Blinking dynamics is usually 
quantified with an exponent that describes the power law sojourn time (see details below).
This characterization is obtained by thresholding the data in order to distinguish between ``on'' and ``off'' states. 
However, thresholding is sometimes scrutinized since the threshold value is rather arbitrary 
and hence a power spectral analysis is postulated to be a preferred tool 
\cite{Pelton2004,PeltonNanoLett2010,Frantsuzov}. Power spectrum is of course the most typical tool used to quantify 
noise.  

Quantum dot (QD) intermittency has attracted considerable 
attention due to the intriguing optical properties
of zero-dimensional materials as well as the power law statistics 
of ``on'' and ``off'' times \cite{Nirmal,Frantsuzov2008}.
Due to the scale-free properties of power law statistics, intermittency naturally yields a power spectrum 
of the form $1/f^{\beta}$ \cite{Manneville}. 
In this report we measure the power spectra of blinking QDs, 
namely we investigate individual nanoscale systems avoiding ensemble averaging. 
We address the fundamental question, whether the standard picture of 
blinking, found also in organic fluorophores, is characterized 
by a single exponent? We show that the description of QD power spectrum with 
a single exponent is incomplete since it hides rich physical phenomena. 
Instead, the power spectrum of these systems is characterized by 
four exponents denoted $\beta$, $z$, $\omega$, and $\gamma$ and, importantly, we explain the physical meaning of these 
critical exponents. The power spectrum of blinking dots turns out to be unusual 
in the sense that it ages with experimental time. Roughly speaking, the longer is the observation time,
the level of noise decreases. 
More specifically, the power spectrum ages as $t^{-z}$ with $z>0$ and $t$ is the measurement time.
In this sense, the power spectrum is non-stationary, in the spirit of Mandelbrot suggestion \cite{Mandelbrot}.  
While the focus of this report is on blinking QDs, 
we believe that the underlying non-stationary behavior
describes a large number of self-similar intermittent systems at the nanoscale, 
including to name a few, liquid crystals \cite{Silvestri}, biorecognition \cite{Bizzarri},
nanoscale electrodes \cite{Krapf}, and organic fluorophores \cite{hoogenboom2007power}. 

\section{Theoretical model}
\begin{figure}
\centerline{\includegraphics[width=10 cm]{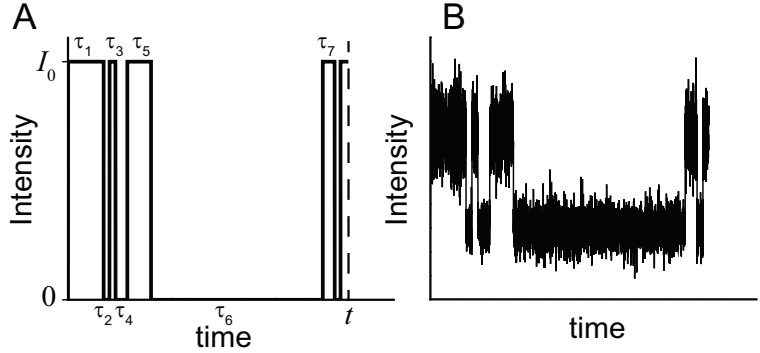}}
\caption{\label{fig:model} 
Simplified model for QD intermittency. 
(A) An individual QD alternates between states ``on'' and ``off'' 
with intensities $I_0$ and zero. 
The sojourn times are $\tau_j$ where $j$ is respectively
odd and even for ``on'' and ``off'' states. 
The measurement time is $t$.  
(B) Additional Gaussian noise in the ``on'' and ``off'' levels is depicted 
so that the intensity in these states is not constant but it fluctuates 
around the mean.}    
\end{figure}

The simplest way to model intermittency is within the assumption of 
a two-state process. 
The QD switches between an active state where the intensity of the signal is $I_0$ 
to a passive state where the intensity is zero. 
This model is sketched in Fig. \ref{fig:model}A.
The sojourn times $\{ \tau \}$ in states ``on'' and ``off'' 
are independent and 
identically distributed (i.i.d.) random variables with, 
for the sake of simplicity, a common 
probability distribution function 
$\psi(\tau)\sim \tau^{-(1+\alpha)}$.  A particularly interesting situation arises when  
$0<\alpha<1$ because then the mean sojourn time diverges and thus the 
process lacks a characteristic time. Otherwise, for $\alpha >1$, the mean sojourn time is finite. 
Diffusion controlled mechanisms of QD blinking lead to $\alpha=1/2$, though measurements 
show deviations from this behavior, suggesting that a wider spectrum of exponents 
$1/2<\alpha<1$ is more suitable. 
For the two-state process, the system yields a power spectrum 
\begin{equation}
S_t(f) \sim {\frac{A_t} {f^\beta} },\label{1/f}
\end{equation}
with the exponent $\beta=2-\alpha$ when $0<\alpha<1$ \cite{Niemann2010,Margolin,Niemann2013,godec2013linear}. 
When the signal is measured over a finite experimental time $t$, 
the power spectral density (PSD) is typically estimated 
using the periodogram method, 
$S_t(f)=\tilde{I}(f,t)\tilde{I}(-f,t)/t$, 
where $\tilde{I}(f,t)$ is the Fourier transform of the 
intensity, $\tilde{I}(f,t)=\int_0 ^t I(\tau) \exp(-i2\pi f\tau) {\rm d} \tau$.
Using this method, it was shown theoretically that the power spectrum decays with 
experimental time \cite{Niemann2013}. The time dependence of the 
spectrum can be found from simple scaling arguments. Because both states are assumed 
to be identically distributed, the total power is a constant. Thus, 
\begin{equation}
\int_{1/t} ^\infty S_t(f) {\rm d} f = A_t \frac{\left( {1/t}\right)^{-\beta + 1}}{|\beta-1|} = \mbox{const},
\label{eqZ}
\end{equation}
where integration is performed from $1/t$, which is the lowest measured frequency. Therefore, 
\begin{equation}
A_t\sim t^{ - z} \label{aging}
\end{equation}
with $z=\beta-1$. Given the relation $\beta=2-\alpha$, we have 
\begin{equation}
z=1-\alpha.\label{z}
\end{equation}
The exponent $z$ is termed the aging exponent. 
The time dependence of the power spectrum reflects the non-stationarity of the process 
as it indicates that the longer the observation time, the smaller the amplitude 
of the power spectrum. In other words, the longer one observes the system, the QD gets trapped 
in longer and longer dark or bright states, thus the switching rate is effectively reduced and 
$1/f$ noise goes down. To the best of our knowledge, 
these predictions were not yet experimentally tested. 
This simple theoretical model bears non-negligible limitations in the analysis of QD intermittency. 
First, the system is assumed to consist of two identical states. 
Second, noise in the experimental system, beyond the switching events, is neglected (Fig. \ref{fig:model}B).
As we shall see in our results, these simplifications fail to capture some of the observed physics. 
However, these theoretical predictions are an excellent starting point in the analysis of blinking power spectrum. 

\section{Experimental methods: quantum dot imaging} 
Core-shell CdSe-ZnS quantum dots were purchased from Life Technologies (Qdot 655, Invitrogen).
In order to avoid aggregation, the QDs were dispersed in a 1\% (w/v)
bovine serum albumin solution to a final concentration of 1 nM. A 20-$\mu \mathrm{L}$ drop of this
solution was placed on a glass coverslip (Warner Instruments,
Hamden, CT) that had been cleaned by sonication
in acetone and ethanol. After a 10-minute incubation period the coverslip was
thoroughly rinsed with deionized water and dried with nitrogen.
We recorded the fluorescence from 1,200 QDs for 22 min
at room temperature in a
Nikon Eclipse Ti TIRF/widefield fluorescence microscope. QDs were excited by a 488-nm laser line
and the emission was collected with a bandpass
filter. Images were acquired in a
frame-transfer electron multiplying charge-coupled device
(EMCCD iXon DU-897, Andor, Belfast UK) at 50 frames per second
(exposure time of 20 ms). 

QD intensities were measured using an automated algorithm implemented
in LabView, which computes the total intensity of each QD.
Due to spatial inhomogeneities in excitation power and dot-to-dot variations
in quantum yield, different QDs can have varying fluorescence
intensities. Thus, we normalize the data so that all intensities lie between
zero and one, allowing us to work with a more convenient
dimensionless intensity system.
For this purpose,
we first subtract the minimum value of the intensity along each QD trace
and then divide it by the maximum (after subtraction) intensity value.

\section{Results}
\begin{figure}
\centerline{\includegraphics[width=11 cm]{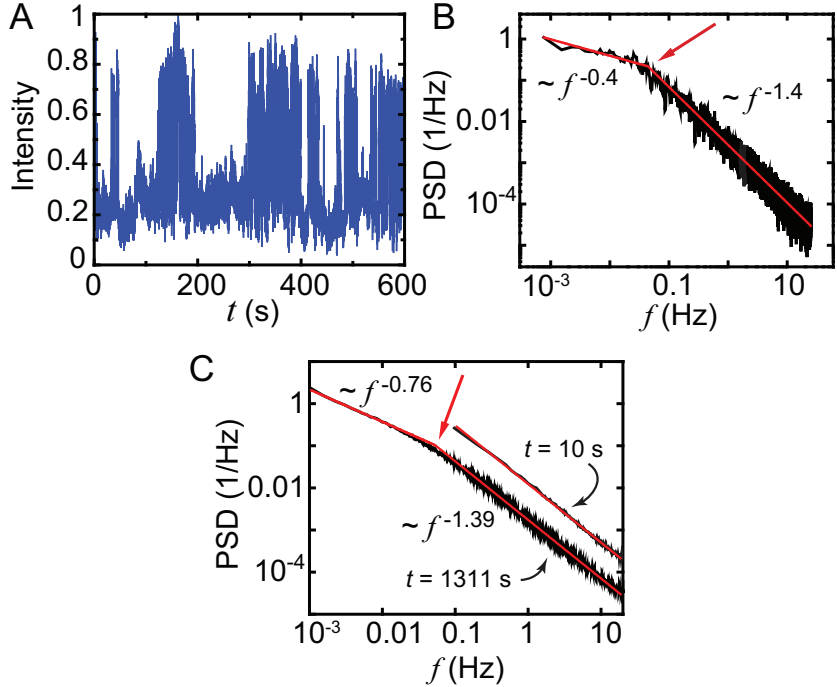}}
\caption{\label{fig:fig1} 
Intermittency in QD fluorescence. 
(A) Normalized fluorescence intensity of an individual CdSe-ZnS QD, 
i.e., maximum intensity is unity. 
(B) PSD of the emission from a single QD measured for 1311 s. 
(C) Average PSD from the emission of $1,200$ individual QDs. The experimental 
times are 10 and 1311 s. The short-time spectrum
is shifted for clarity. The lines show linear regression of the 
log-log plot for high and low frequencies according to Eq.~(\ref{1/f}), 
and the red arrows point to the transition frequency $f_{T}$.}    
\end{figure}

Figure \ref{fig:fig1}A shows the first $600$ seconds of the normalized intensity trace of a typical QD. 
Usually, QD blinking is analyzed by using a threshold that defines bright and dark 
states \cite{Kuno,Shimizu,Brokmann}.  
As mentioned, the threshold determination is rather arbitrary, 
and in that sense power spectrum analysis is preferred \cite{PeltonNanoLett2010}. 
In agreement with previous observations, 
the distribution of ``off'' times is 
well described by a power law $\psi_{\mathrm{off}}(\tau)\sim \tau^{-(1+\alpha)}$; 
whereas, the distribution of ``on'' times shows truncated power-law 
behavior $\psi_{\mathrm{on}}(\tau)\sim \tau^{-(1+\alpha)}e^{-\tau/\tau_{\mathrm{on}}}$ \cite{Shimizu}. 
In our data we find $\alpha=0.63\pm0.10$ and $\tau_{\mathrm{on}}=8.5$ s, 
a time scale that will soon become important.

A representative power spectrum from an individual QD is shown in Fig. \ref{fig:fig1}B. 
The experimental time of the time trace employed in the computation of this spectrum is 1311 s, 
i.e., the whole available time. 
Since the normalized intensity is dimensionless, the PSD has units of $\mathrm{Hz}^{-1}$. 
Figure \ref{fig:fig1}C 
shows power spectral densities obtained from averaging the spectra 
of $1,200$ individual QDs for experimental times of 10 and 1311 s. 
The spectrum of the long-time trace 
exhibits two regimes with distinctive $1/f^\beta$ behavior. 
For frequencies below a transition frequency $f_{T}$ we have $S_t(f) \sim
f^{ - \beta_{<} }$ with $\beta_{<}=0.76 \pm 0.02$ 
($n=1,200$ traces)
while above this frequency 
we have $S_t(f) \sim f^{- \beta_{>} } $ with $\beta_{>} = 1.393\pm 0.002$.
The separation into two 
regimes is caused by the cutoff that characterizes the ``on''-time 
distribution. Hence, the transition frequency $f_{T}$ is, not surprisingly, 
of the order of $1/ \tau_{\mathrm{on}}$. We will soon compare 
our experimental findings with the theory 
for $\beta_{>}$ and $\beta_{<}$, but let us first discuss the transition frequency.

\subsection{The transition frequency $f_{T}$ }
The existence of a transition frequency implies that the measurement
time is crucial. The PSD of 
the short-time trace (Fig. \ref{fig:fig1}C with $t=10$ s) displays $1/f^{\beta}$ spectrum 
with a single spectral exponent $\beta_{>}$. 
On the other hand, long enough measurements yield the 
transition to a different behavior.
Importantly, an observer 
analyzing short-time traces would reach the conclusion that
the power spectrum is non-integrable, since $\beta_{>} > 1$.
If we wait long enough we 
eventually observe integrable $1/f$ noise, 
since $\beta_{<} < 1$. 
In some sense we are lucky to observe this transition: it is detected
since  the cutoff time $\tau_{\mathrm{on}}$ is on a reasonable time scale. 

Previously, Pelton et al. have reported a transition frequency in the power spectrum 
of blinking QDs \cite{Pelton2007}. However, that transition has a different nature from 
the one reported here. In our measurements, a cutoff in ``on'' sojourn times introduces 
a transition from $1/f^{2-\alpha}$ to $1/f^{\alpha}$ at low frequencies, i.e., long time behavior, 
with $f_T$ of the order of 0.06 Hz. 
On the other hand, Pelton et al. find a high frequency transition, i.e., short time behavior, where the spectrum 
shifts from $1/f^{2-\alpha}$ to $1/f^2$ at frequencies above the transition. 
This high frequency transition was found to be of the order of 100 Hz. 
The transition to $1/f^2$ spectrum was interpreted as 
short time carrier diffusion yielding, at high frequencies, the power spectrum characteristic of Brownian motion \cite{Pelton2007}. 

\subsection{Spectral exponents $\beta_{<}$ and $\beta_{>}$ }
Both exponents $\beta_{<}$ and $\beta_{>}$ are related to $\alpha$.
The exponents measured in this study are reported in Table 1 
for the benefit of the reader.
For frequencies $f>f_{T}$, the cutoff time is of
no evident relevance and both states are effectively
distributed with power law statistics $\psi(\tau) \sim \tau^{- (1 + \alpha)}$.
In this regime, theory 
predicts $\beta_{>} = 2- \alpha$ as mentioned above. Since $\alpha=0.63\pm0.10$, 
we expect $\beta_{>} = 1.37\pm0.10$, which is similar to the measured exponent.
In contrast, for $f<f_{T}$ we must consider the effect of the
cutoff time. 
Thus, we define a modified model, which includes a cutoff 
in the distribution of ``on'' times, so that the 
probability density functions of ``on'' and ``off'' sojourn times are different. 
The important feature of this model
is that the mean ``on'' sojourn time is finite, 
which modifies the underlying exponents that describe the power spectrum.
For this case, 
one finds $\beta_{<} = \alpha$. Experimentally we find $\beta_{<}=0.76\pm 0.02$ while $\alpha=0.63\pm 0.10$, 
so small deviations are found. The theoretical sum rule
$\beta_{<} + \beta_{>} = 2$ is insensitive to the value
of $\alpha$ provided that  $\alpha<1$, since this implies 
the divergence of the mean ``off'' sojourn time, which is the main condition
for the observed self-similar behavior.    

\subsection{Aging exponent $z$} 
\begin{figure}
\centerline{\includegraphics[width=11 cm]{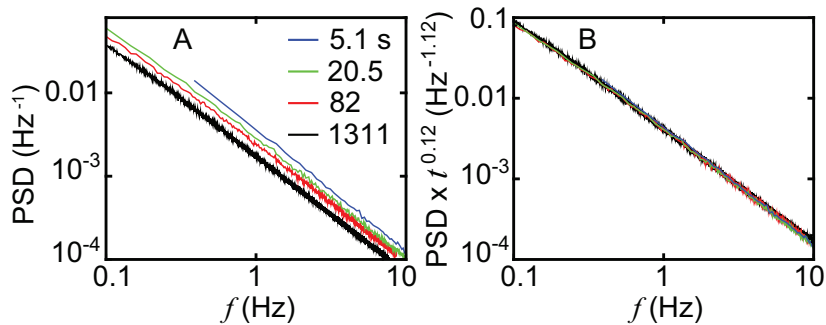}}
\caption{\label{fig:fig2}  
Aging of the power spectrum. 
(A) Average of 1200 QD power spectral densities for four different experimental 
times: $5.1$, $20.5$, $82$, and $1311$ s. 
(B) When the power spectral density is multiplied by an aging factor $t^z$ 
with $z=0.12$, 
where $t$ is the experimental time, the spectra collapse to a single trace.}
\end{figure}

Since the ``off'' sojourn times are scale free, i.e., mean time is infinite, 
we expect the amplitude of the power spectrum $A_t$ to depend
on measurement time. 
Figure \ref{fig:fig2}A shows averaged power spectra computed for trajectory lengths 
of $5.1$, $20.5$, $82$, and $1311$ s. As the experimental time $t$ increases, the 
magnitude of the PSD is not constant but it decreases.
To the best of our knowledge, this is the first experimental report explicitly showing 
that $1/f$ noise in nanoscale systems ages and the concept of stationary
$1/f$ noise, so popular in a vast literature, breaks down. 
Figure \ref{fig:fig2}B shows that the PSD data collapse 
to a single master trace when multiplied by $t^{0.12}$.
According to the theory \cite{Niemann2013}, 
the power spectrum amplitude scales as $A_t\sim t^{-z}$ with the exponent $z=1-\alpha$ 
both below and above the transition frequency. 
Thus we expect $z=0.37\pm0.10$, 
which is slightly larger than the measured value of the aging exponent $z =0.12$. 
We will address this deviation with simulations showing that additional noise
in the ``on'' state (see Fig. \ref{fig:model}) is important.

\subsection{The zero frequency exponent $\omega$}
\begin{figure}
\centerline{\includegraphics[width=11 cm]{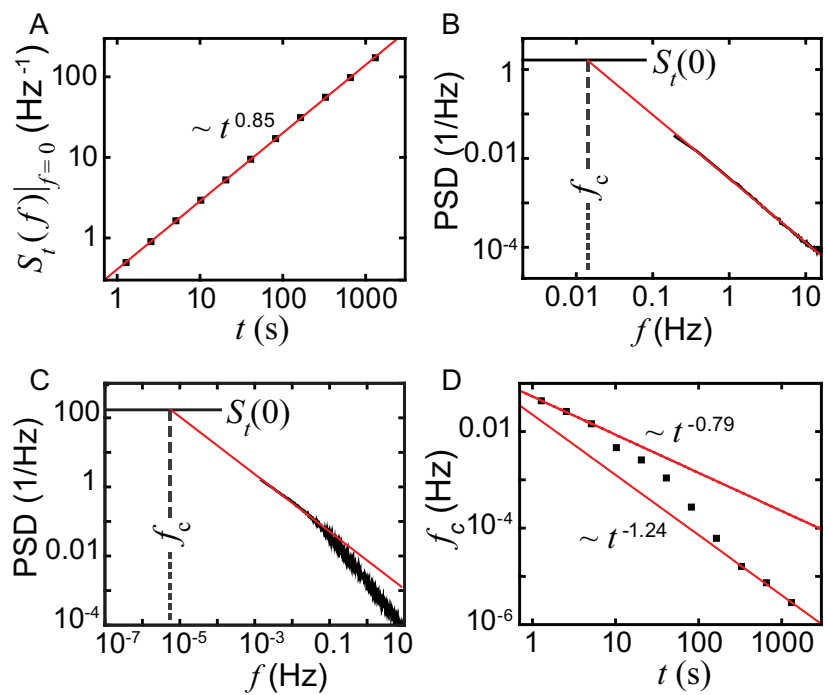}}
\caption{\label{fig:fig3} Additional critical 
exponents describing QD intermittency. 
(A) Zero-frequency spectrum vs. experimental time.
(B-C) Examples showing how the crossover frequency $f_c$
is found from the average power spectrum. The horizontal line shows 
$S_t(0)=S_t(f)|_{f=0}$. The crossover frequency $f_c$ is found by extrapolating  
$S_t(f) \sim { A_t f^{-\beta} }$ to the intersection with $S_t(0)$. 
In b the experimental time is $t=5.1$ s, thus $S_t(f) \sim { A_t f^{-\beta_{>}} }$. 
On the other hand, in c the time is $t=1311$ s, thus the spectrum shows 
two different frequency regimes, with $S_t(f) \sim { A_t f^{-\beta_{<}} }$ 
for $f<f_T$. Note that $f_c$ shifts by more than three orders of magnitude 
between 5.1 and 1311 s.
(D) Crossover frequency vs. experimental time. 
We find that $S_t(0)\sim t^\omega$ with $\omega=0.85$ and, for short 
times, $f_c\sim t^{-\gamma}$ with $\gamma=-0.79$.}
\end{figure}

Next, we define the spectrum at zero frequency with
\begin{equation}
S_t(0)=S_t(f)|_{f=0}=\frac{(\int^t_0 I(\tau)d\tau)^2}{t},
\label{eqEE}
\end{equation}
and the corresponding exponent 
$S_t(f)|_{f=0} \sim t^{\omega}$. 
Notice that $S_t(0)/t$ is merely the square
of the time average $\int_0 ^t I(\tau) {\rm d} \tau /t$ and
its experimental evaluation does not require a fast Fourier transform.
For stationary and ergodic processes with non-zero mean intensity
we have normal behavior $\omega=1$.  
On the other hand, if $\omega<1$ the average intensity decays to zero. 
As shown in Fig. \ref{fig:fig3}A, our measurements yield $\omega\simeq 0.85$, 
which is a second indication of non-stationarity. 

In our system, for long experimental times, the 
``on'' time distribution displays a cutoff and thus 
the mean ``on'' time is finite. Therefore, the expected area 
under the intensity time trace can be estimated to be 
\begin{equation}
\langle \int_0 ^t I(\tau) {\rm d} \tau \rangle= \langle n \rangle \tau_{\mathrm{on}} I_{\mathrm{on}},
\label{integralI}
\end{equation}
where $I_{\mathrm{on}}$ is the intensity in the ``on'' level 
and $\langle n \rangle$ is the average number of 
renewals up to time $t$, i.e., the number of switchings, 
which is known to increase as $t^\alpha$ for waiting ``off'' times 
distributed according to power laws \cite{GodrecheLuckJSP,Klafter}. 
Hence we see that theoretically $S_t(f)|_{f=0} \sim t^{2 \alpha -1}\simeq t^{0.26}$. 
Within this model, $\omega=2 \alpha -1=0.26$. 
The measurements in Fig. \ref{fig:fig3}A give $\omega=0.85$,
which is surprising since we expect that, at least for long times 
compared with $\tau_{\rm{on}}$, 
the cutoff in the ``on'' times dictate the behavior of the zero frequency spectrum. 
We will soon remove this mystery by detailed consideration of 
the effects of noise in the ``on'' and ``off'' states 
using numerical simulations. 
What becomes clear is that the standard description of blinking 
systems with a single exponent $\alpha$, 
so popular in the literature does not describe aging 
accurately and needs to be expanded. 
Namely, in our measurements, $\omega$ is not obtained 
from $\alpha$ in a straightforward way. Hence the standard 
picture of these systems is challenged.

\subsection{The crossover frequency $f_c\sim t^{- \gamma}$}
The transition between the zero frequency spectrum $S_t(0)$ and the small but finite frequency
behavior $S_t(f) \sim f^{-\beta}$, defines a crossover or cutoff frequency 
$f_c$. A crossover frequency is many times assumed to be time
independent, though its observation may require extremely long measurement times. 
For example, in spin glasses the inverse of the cutoff frequency was estimated to be of the 
order of age of universe \cite{Kogan}, and hence it cannot be directly investigated.  
Given that the PSD ages, we hypothesize the cuttoff frequency also 
changes with experimental time. 
We investigate the time dependence of $f_c$ within our observation
window, which is long in the sense that we measure thousands of 
transitions between ``on'' and ``off'' states. In order to estimate $f_c$, we extrapolate 
Eq.~(\ref{1/f}) to the intersection with the zero-frequency spectrum, 
given by Eq.~(\ref{eqEE}), as shown in Figs. \ref{fig:fig3}B and \ref{fig:fig3}C. 
According to the two state model with i.i.d. sojourn times, 
$f_c \sim t^{-\gamma}$, and $\gamma=1$ \cite{Niemann2013}. 
Again this behavior can be derived using scaling arguments. 
First, we use idealized models, which, by noting that at the crossover frequency 
$A_t f_c^{-\beta}=S_t(0)$, give 
$t^{-z} f_c^{-(2-\alpha)} \sim t$ for short times and 
$t^{-z} f_c^{-\alpha} \sim t^{2\alpha-1}$ for long times. 
Surprisingly, these scaling arguments predict 
$f_c \sim 1/t$ for all times, independent of $\tau_{\rm{on}}$. 
Hence, in this case $\gamma=1$. However, since we already observed 
deviations in $\omega$ and $z$ from the idealized two-state model, 
we now use a more general approach using scaling arguments. 
Here, a second scaling approach relates the various exponents 
that characterize the process. As before, we employ the relation 
$A_t f_c^{-\beta}=S_t(0)$, which yields $t^{-z} f_c^{-\beta}\sim t^\omega$ 
from the definitions of the various critical exponents. 
Thus we find 
\begin{equation}
\gamma=\frac{\omega+z} {\beta}.
\label{gamma}
\end{equation}
Using measured values for $\omega$, $z$, and $\beta$, we have $\gamma=0.70$ for short times 
(see Table 1, $\omega=0.85$, $z=0.12$, and $\beta_>=1.39$) and 
$\gamma=1.27$ for long times ($\beta_<=0.76$).

For short times, we observe in experiments 
that the crossover frequency scales with experimental time 
as $f_c \sim t^{-0.79}$ (Fig. \ref{fig:fig3}D). Thus $\gamma=0.79$, which is in good 
agreement with our general scaling argument approach (Eq. \ref{gamma}) and is consistent with 
the other critical exponent measurements. 
For longer measurement times, as seen in Fig. \ref{fig:fig1}C, the low frequency 
spectrum shifts from $S_t(f)\sim 1/f^{\beta_{>}}$ to $S_t(f)\sim 1/f^{\alpha}$ 
where $\beta_{>}>\alpha$. As a consequence of this effect, a transition 
is observed roughly on the cutoff time
$\tau_{\rm{on}}$  and $f_c$ decays faster, 
namely, $\gamma>1$ for $t>\tau_{\rm{on}}$. 
The behavior can be qualitatively understood, by comparing Figs. \ref{fig:fig3}B and \ref{fig:fig3}C. 
As the slope of the power spectrum becomes less steep, the crossover is rapidly shifted to smaller 
frequencies. The value for $\gamma$ at long times is difficult to estimate from our measurements, 
but it is roughly $\gamma=1.24$. Again, this value is consistent with the measured values of $\omega$ and $z$ as 
predicted by scaling arguments (Eq. \ref{gamma}).

\subsection{Numerical simulations}
Deviations between experiments and theory 
can arise from at least three sources: experimental noise, finite 
measurement time, and model assumptions not being realistic. 
Recall that the models neglect any physical noise beyond the switching events 
(see Fig. \ref{fig:model}A) 
and only attempt to solve for convergences in the long time limits. 
In particular, the intensity in the ``on'' and ``off'' states are not constant; 
the signal is always fluctuating.
To address this issue we turn to numerical simulations. 
We estimated the four exponents $\beta$, $\omega$, $\gamma$, and $z$ 
based on numerical simulations. In simulations we add noise to the 0/1 
signal (idealized model). Hence, simulations provide 
additional insight on the analysis. 
The performed simulations are:
PL: Power law; 
PLN: Power law with noise; 
PLC: Power law with cutoff (truncated ``on'' times); and
PLCN: Power law with cutoff and noise.  

Initially, we generated time series of on/off states with random waiting times drawn from a
power-law distribution $\psi(\tau)= \alpha t_0^\alpha/(\tau+t_0)^{-(1+\alpha)}$. The constant
$t_0$ was chosen to be equivalent to the experimental binning time, $t_0=20$ ms, and $\alpha=0.63$.
We refer to this simulation as PL. 
In order to add Gaussian noise to the realizations (PLN), the ``on'' and ``off'' intensities 
were transformed at each sampling time into normal random variables $N(0.7, 0.04)$ and
$N(0.2, 0.0064)$, respectively. The sampling time was chosen to be $20$ ms. 
The variance difference reflects the increased level of 
noise in the ``on'' state due to shot noise. 
To simulate sojourn times distributed according to a power law with cutoff (PLC), the ``on'' times were drawn 
from a distribution $\psi(\tau)\sim \tau^{-(1+\alpha)} \mathrm{exp}(-\tau/\tau_{\mathrm{on}})$ and $\tau\ge t_0$. 
The cutoff time was chosen to be $\tau_{\mathrm{on}}=15$ s. Additionally, 
we performed simulations with both cutoff ``on'' times and added noise (PLCN). 
Once noise is added the sojourn time distributions change and estimations of 
$\alpha$ and $\tau_{\mathrm{on}}$ generally shift toward lower values. Therefore we chose 
$\tau_{\mathrm{on}}=15$ s in our simulations instead of $8.5$ s as measured in experiments. 

%
%
The combination of a cutoff time and Gaussian noise has significant effects 
on the zero-frequency spectrum $S_t(0)$ and the crossover frequency $f_c$ (Fig. S1).
In these simulations, $f_c\sim t^{-\gamma}$ with $\gamma<1$ at short times and 
$\gamma>1$ at times $t>\tau_{\mathrm{on}}$, as observed in the experimental data, and 
the zero frequency spectrum scales as $S_t(0)\sim t^{0.84}$ as well. 
Table 1 summarizes the 
exponents found in both experimental data and numerical simulations. 
Simulation results from a power law distributed two-state model with i.i.d. sojourn times and without noise (PL) 
are also shown in the Table along with this model's theoretical predictions.
We observe that, while basic non-stationary features of $1/f$ noise agree with recent theory \cite{Niemann2013}, 
our experimental work shows that introducing noise in the ``on''/``off'' levels and
a finite mean ``on'' sojourn time is crucial for a complete 
picture of the power spectrum of blinking QDs. 

 \begin{table}
\caption{\label{tab:table1}
Exponents that describe $1/f$ noise for QD emission (experimental data) and simulations. 
Results from numerical simulations of two dichotomous random processes 
are shown: power-law distributed waiting times (PL) and power law with cutoff in ``on" times 
and noise (PLCN). 
The exponents describe the power spectrum, aging, crossover frequency, 
and zero-frequency spectrum.}
\begin{indented}
\lineup
\item[]\begin{tabular}{@{}clcccc}
\br
  &&&Experimental data&\multicolumn{2}{c}{Numerical simulations}\\
 &&Theory \cite{Niemann2013}&QD&PL&PLCN\\ 
\mr
 $\beta_{>}$	& $S_t(f)\sim1/f^\beta$	&$\beta_{>}=2-\alpha$	&1.39&1.38&1.35\\
 $\beta_{<}$	&						&$\beta_{<}=\alpha$		&0.76&&0.62\\
 $z$			& $A_t\sim 1/t^z$		&$z=1-\alpha$			&0.12&0.36&0.31 \\
 $\gamma$		& $f_c\sim 1/t^\gamma$	&$\gamma=1$				&0.79$^{\rm a}$&0.99&0.72$^{\rm a}$\\
 $\omega$		& $S_t(0)\sim t^\omega$	&$\omega=1^{\rm b}$&0.85&0.99&0.84\\
\br
\end{tabular}
\item[] $^{\rm a}$ These results hold for $t<\tau_{\mathrm{on}}$. For longer times $\gamma>1$.
\item[] $^{\rm b}$For long times we have $\omega=2 \alpha-1$.
\end{indented}

\end{table}

As seen in Table 1, the addition of noise and cutoff on the ``on''-time distribution 
modifies the exponents in such a way that now we obtain better agreement between simulations 
and measurements. Jeon et al. discussed theoretically the strong influence 
of noise on the evaluation of physical parameters from data exhibiting 
power law distributed sojourn times \cite{JeonNoiseCTRW}. While that work focused on diffusion 
of individual molecules in living cells, we can infer the relevance of noise 
also in our blinking system. Roughly speaking, when power law sojourn times 
are so broad, the system remains in a state (e.g, ``off'') for a time 
that is of the order of the measurement time. Therefore, the noise level 
in this long state is of utmost importance for a detailed analysis. 

\section{Discussion}
Our data show that the power spectrum of blinking quantum dots
crucially depends on measurement time. We present the first experimental
evidence for Mandelbrot's suggestion that $1/f$ noise is related to non-stationary
signals. Our measurements were performed at the nanoscale by measuring single particles, 
thus removing the problem of averaging a large number of particles, 
typically found in macroscopic systems. 
The most common description of macroscopic $1/f$ electronic noise, 
commonly referred to as the McWhorter model \cite{Balandin,Kogan}, stems from the observation that 
a superposition of Lorentzian spectra with a broad distribution of relaxation times 
yield $1/f$ noise. If this philosophy would hold at the nanoscale, by probing 
an individual molecule, one would expect to measure a Lorentzian 
spectrum with a well-defined relaxation time. This scenario is not found for blinking quantum dots. Instead, 
$1/f$ is observed at the nanoscale. Further, the noise exhibits clear 
non-stationary behavior, i.e., dependence on measurement time. 
We quantify this non-stationarity with critical
exponents. In particular, the aging exponent $z$ shows that the amplitude of the noise decreases
as a power law with time. This effect should also be found in other intermittent systems.

The key finding in our work shows that $1/f$ power spectrum of intermittent QDs
decays with experimental time, i.e., it ages, and thus the spectrum does not converge in long time measurements 
as typically assumed for standard stochastic processes. 
These results agree with previous observations that analyzed blinking in semiconductor 
quantum dots as a non-stationary process \cite{Brokmann,messin2001bunching}.
The description of non-stationary $1/f^\beta$ noise we present is vastly different from traditional
approaches that characterize it with a single exponent. 
Besides $\beta$, three additional exponents 
give the dependence on the measurement time.
These exponents describe aging of the power spectrum $A_t \sim t^{-z}$, 
the zero-frequency spectrum $S(f)|_{f=0} \sim t^\omega$, and the low cutoff 
frequency $f_c\sim t^{-\gamma}$. 
Importantly, the appearance of a transition frequency due to a 
finite mean ``on'' sojourn time, modifies the underlying exponents that describe the power spectrum.

In an observation time $t$, the total power of the process
is $\int_{1/t} ^\infty S_t(f) {\rm d} f$ where $1/t$ is the 
lowest measured frequency. For a process with 
power spectrum $S(f)=A/f^\beta$ with $1<\beta<2$ the total power
diverges as time increases, 
due to the low frequency behavior. In contrast, if
$S(f) = A/ f^\beta$ and $0<\beta<1$, the total power diverges due to the high frequency
behavior of the spectrum. We observe that in QDs, two different phenomena 
limit the increase of total power. First, below the transition frequency $f_T$, 
we find $\beta<1$ due to the cutoff in the ``on'' sojourn times and hence the spectrum is integrable 
at low frequencies. For large frequencies $\beta > 1$ hence it is integrable also at 
high frequencies. Second, 
as the observation time progresses, the amplitude of the power spectrum 
decreases with $A_t \sim t^{-z}$, so that $S_t(f)\rightarrow 0$
in the limit $t\rightarrow \infty$. Both these findings maintain the
total power finite. 
More precisely, if one measures for times that are shorter than 
$\tau_{\mathrm{on}}$ and hence the transition to 
an integrable spectrum is not detected at low frequencies, the decrease of the spectrum with time 
ensures that the total area under the power spectrum does not blow up. 
This will become particularly important in the limit of weak laser field excitation 
and low temperatures, where $\tau_{\mathrm{on}}$ becomes extremely large \cite{Shimizu} and 
a single regime in the power spectrum holds for all observable time scales, 
i.e., the transition frequency $f_T$ is not observed within the available frequency range.
We measure the total power by integrating the
power spectral density as defined above
and indeed, we obtain a finite value which is not diverging, i.e. it is bound, 
though convergence is slow (Fig. S2).
The aging of the spectrum ensures that total power of the system 
will not diverge, hence, our observations 
help in removal of a long-standing paradox of physics \cite{Mandelbrot}.

By resorting to numerical simulations, 
we find that a model that includes both a cutoff in ``on'' sojourn times and noise in 
each state (PLCN model) describes more accurately the experimental results obtained. 
These simulations emphasize the influence of noise within the ``on'' and ``off'' states and power low with a cut-off distribution. 
However, the value of the aging exponent $z$ still remains somewhat far from our experimental observations. 
The aging exponent in simulations is estimated to be $z=0.31$, while in experiments $z=0.12$.
We can speculate there are two reasons responsible for the observed discrepancy. 
First, these models still assume the existence of solely two levels. 
Recent experiments point to the existence of intermediate states in the emission 
from core-shell QDs \cite{zhang2006continuous,spinicelli2009bright}. The occurrence of multiple states has been 
described both in terms of blinking processes that are faster than the time resolution of the experiment 
(as is the case in our experiments) \cite{Galland} 
and in terms of multiple physical states within the core-shell quantum dot 
\cite{spinicelli2009bright,amecke2011intermediate}. 
Nevertheless, the main aspects of the non-stationarity described here are expected to hold for the 
multiple level system as long as at least one of the states is governed by a 
scale-free power law distribution. Second, a different phenomenon that could affect the 
measured critical exponents is noise in the QD levels that is not Gaussian. 
The influence of non-Gaussian noises can of course have striking consequences in 
the properties of a stochastic process. 

In our approach the exponents were analyzed in a way that is reminiscent of critical behavior,
with scaling relations showing the exponents are dependent on each other 
and with different behaviors below and above a transition frequency $f_T\simeq 1 / \tau_{\mathrm{on}}$. 
These results are relevant to a broad range of systems
displaying power law intermittency \cite{Zumofen,Scher,Silvestri,Bizzarri,Krapf}. 
Further, power law sojourn times, which is the
basic ingredient leading to the observed non-stationary spectrum in blinking quantum dots, 
are widespread and are found in glassy systems \cite{Bouchaud,burov2010aging} 
and anomalous diffusion in live cell environments and
other complex systems 
\cite{bouchaud-georges,metzler-klafter,weigel2011PNAS,tabei2013intracellular,burov2013distribution}. 
Therefore, the measured exponents could be a general feature of many 
noisy signals. 
Finally, the traditional characterization of blinking quantum dots with a single exponent $\alpha$ 
is shown to be limited
and to hide interesting physics described by different critical exponents. 

\section{Conclusions}
Our experiments show how the analysis of noise in blinking
quantum dots reveals rich physical behavior described by four critical
exponents.  This is vastly different from traditional
approaches that characterized power spectrum of $1/f^\beta$ noise with a
single exponent $\beta$. The exponent $z$ describes the aging of the spectrum with
measurement time, showing a decrease of the noise level as
the measurement time increases. The exponent $\beta$ describes the 
$1/f^\beta$ noise as in many previous studies, however we find two
such exponents $\beta_{<}$ and $\beta_>$, below and above the transition frequency $f_T$. 
The zero frequency exponent $\omega$ describes
the time average of the intensity, which is essentially related to ergodicity
and yields further information on the non-stationarity of the process. 
The exponent $\gamma$ describes the crossover from zero frequency to $1/f^\beta$. 
We hope that our work will promote measurements of exponents of $1/f^\beta$
spectrum, since they reveal the true complexity of the observed
phenomena. 

\ack{SS thanks Elizabeth Akin, Philip Fox, and Michael Tamkun for experimental help. 
EB was supported by the Israel Science Foundation. 
DK acknowledges the support from the National Science 
Foundation under grant 1401432.}

\clearpage

\bibliographystyle{iopart-num}
\bibliography{apssamp} 



\end{document}